# REVIEW AND ANALYSIS OF THE ISSUES OF UNIFIED MODELING LANGUAGE FOR VISUALIZING, SPECIFYING, CONSTRUCTING AND DOCUMENTING THE ARTIFACTS OF A SOFTWARE-INTENSIVE SYSTEM

Dr.S.S.Riaz Ahamed

Principal, Sathak Institute of Technology, Ramanathapuram,India.

Email:ssriaz@ieee.org, ssriaz@yahoo.com

**ABSTRACT**

The UML allows us to specify models in a precise, complete and unambiguous manner. In particular, the UML addresses the specification of all important decisions regarding analysis, design and implementation. Although UML is not a visual programming language, its models can be directly connected to a vast variety of programming languages. This enables a dual approach to software development: the developer has a choice as to the means of input. UML can be used directly, from which code can be generated; or on the other hand, that which is best expressed as text can be entered into the program as code. In an ideal world, the UML tool will be able to reverse-engineer any direct changes to code and the UML representations will be kept in sync with the code. However, without human intervention this is not always possible. There are certain elements of information that are lost when moving from models to code. Even then, there are certain aspects of programming language code do seem to preserve more of their semantics and therefore permits automatic reverse-engineering of code back to a subset of the UML models.

## INTRODUCTION

The UML is a language in that it specifies a vocabulary and rules for combining words in that vocabulary. It is a modeling language because its vocabulary and rules focus on the conceptual and physical representation of a system. One of the purposes of UML is to help us visualize software. It enables developers to share what they visualize in their minds or in what would be very proprietary diagrams. Its graphical nature allows us to represent structures that transcend that which can be directly expressed in a programming language, no matter how expressive it may be. And finally, it enables us to explicitly capture those models that are present in the minds of the developers, be it for future reference or plain complementary documentation.

There are four kinds of things in the UML.

*Structural Things*

These are the nouns of the UML models, mostly static, representing elements that are either conceptual or physical. These are:

- **Classes** and **interfaces**. Variations of classes also exist: actors, signals and utilities.
- Collaboration defines an interaction and is a society of roles and other elements that work together to provide some cooperative behavior that is bigger than the sum of all the elements. They have, therefore, structural and behavioral dimensions which can be represented in class and interaction diagrams respectively.
- **Use cases**, which describe a set of sequences of actions that a system performs within an interaction with one or more actors, yielding an observable result of value to one or more of the actors. A use case is realized by collaboration. This step must be done with utmost care since the developer with a prejudice to see things through the algorithmic model can perhaps more easily decompose the various use cases into procedures than step into the new perspective and search for a minimal set of objects, each with well defined capabilities, whose collaboration is able to yield the required functionality.
- **Active classes** are classes whose instances contain one or more processes or threads and therefore have an initiating or controlling role. Its instances imply concurrent execution with other instances. Active classes are usually either a process or a thread.

- A **component** is different from the previous things in that it represents something physical and replaceable in the system – rather than something conceptual. It provides the implementation of a set of interfaces (usually at least one interface) and may depend on interfaces supplied by other components. Components have a wide variety from source files, executables, library (could be a .NET or Java Beans component), database table, ordinary file, ordinary document. A component typically represents the physical packaging of several conceptual things like classes, interfaces and collaborations.
- A **node** is another physical element and it represents a computational resource, usually a computer or device. It may have memory and a processor. It may host a set of components and they may also move from node to node.

*Behavioral Things*

These are the verbs of the models, representing behavior over time and space. These are:

- **Interaction** is a behavior that includes the exchange of a set of messages among a set of objects within a particular context to accomplish a specific purpose. Interactions may specify the behavior of a society of objects or of an individual operation.
- A **state machine** is a behavior that specifies sequences of states an object or an interaction goes through during its lifetime in response to events, together with its responses to those events. The behavior of an individual class or a collaboration of classes may be specified with a state machine.

*Grouping Things*

These are used for organizing things in the UML. There is only one:

- **Packages** are a general purpose mechanism for organizing elements into groups. These may include structural, behavioral and even other grouping things. A package, however, is purely conceptual, existing only at development time, not at run time.

*Annotational Things*

These are the explanatory elements in UML models. They are comments that can be attached to any element in the model, to help describe it or highlight any particular aspect of it. There is only one:

- A **note** is simply a symbol inside of which we may place constraints or comments attached to one or more elements in a model.

**RELATIONSHIPS**

There are four basic relationships in the UML, although there are variations. These are:

- **Dependency** is a semantic relationship between two things in which a change to one thing may affect the semantics of the other thing (the dependent thing).
- **Associations** are structural relationships that describe a set of links, where a link is defined as a connection between objects. Aggregation is a special kind of association that implies a structural relationship between a whole and its parts. In turn, composition is a stronger form of aggregation where the lifetime of the parts is contained within the lifetime of the whole.
- **Generalizations** are specialization/generalization relationships in which the instances of the specialized element are substitutable for instances of the generalized element (the parent).
- **Realizations** are semantic relationships between classifiers, where one classifier specifies a contract that another classifier guarantees to carry out. Realizations occur in two places: between interfaces and the classes or components that realize them, and between use cases and the collaborations that realize them.

**UML SUPPORT FOR ARCHITECTURE**

We should already be convinced that visualizing, specifying, constructing and documenting non-trivial software systems demands that we be able to view the system from a variety of perspectives.

In the creators of the UML propose that "the architecture of a system be described by five interrelated views, where each view is a projection into the organization and structure of the system, focused on a particular aspect of that system".

The static aspects of these views are specific for each view but the dynamic aspects are always described by using interaction diagrams, statechart diagrams, activity diagrams and to a lesser extent with use-case diagrams.

We will detail each of these views by themselves and then explain their interrelationships. You should keep in mind that UML completely supports these views with appropriate diagrams and corresponding semantics.

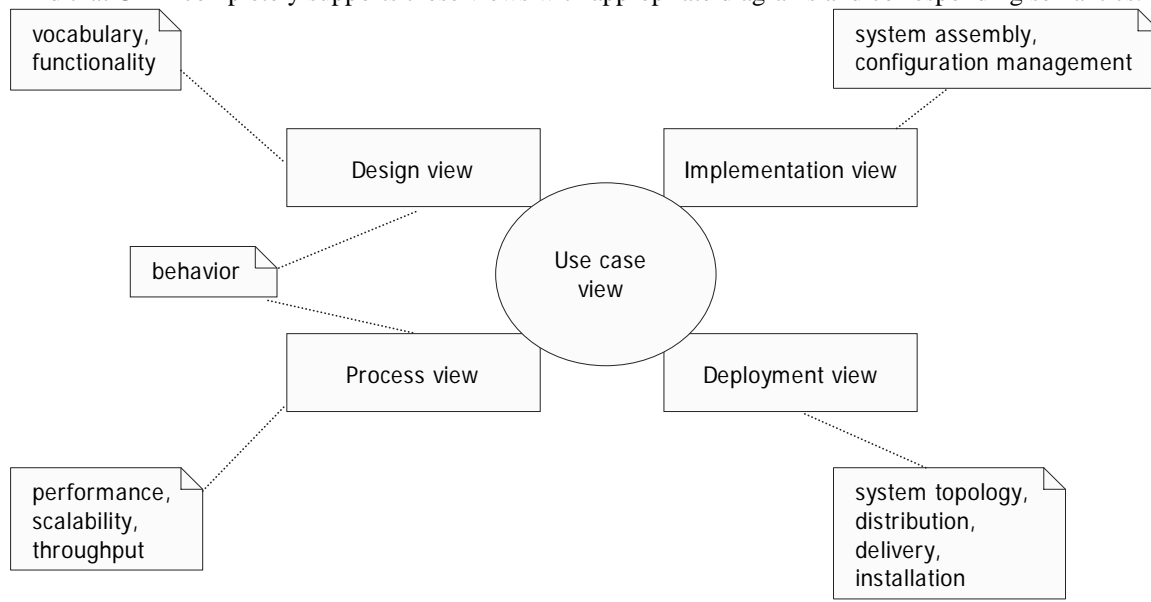

**A system's architecture**

The *use case view* of a system includes the use cases that describe the behavior of the system as seen by its end users, analysts and testers. This view doesn't really specify the organization of a software system. Rather, it exists to specify the forces that shape the system's architecture. The static aspects of this view are captured in use case diagrams.

The *design view* of a system includes the classes, interfaces and collaborations that constitute the vocabulary of the problem and its solution. This view primarily supports the functional requirements of the system, i.e. the services it should provide to its end users. The static aspects of this view are captured in class diagrams and object diagrams.

The *process view* of a system is concerned with the processes and threads that form the system's concurrency and synchronization mechanisms. This view primarily addresses the performance, scalability, and throughput of the system. Using UML, the static and dynamic aspects of this view are captured in the same kinds of diagrams as the design view, but focusing on the active classes that represent its threads and processes.

The *implementation view* of a system deals with the components and files that are used to assemble and release the physical system. This view is primarily concerned with the configuration management of the system's releases, which is composed of somewhat independent components and files that can be assembled in various ways to produce an executable system. The static aspects are represented in component diagrams.

Finally, the *deployment view* of a system includes the nodes that form the hardware topology on which the system will execute. This is the place where hardware meets software. The view primarily addresses the distribution, delivery, and installation of the parts that make up the physical system. This view is very relevant in today's increasingly distributed systems. The static aspects of this view are captured in deployment diagrams.

Each of these five views can be examined independently as needed by the observer. However, they have a very tight integration among them: nodes in the deployment view hold components in the implementation view that, in turn, represent the physical realization of the classes, interfaces, collaborations and active classes from the design and process views.

As explained before, the purpose of the use case view is not to specify the organization of the software system, but rather to specify the forces that have driven the resulting architecture.

**DIAGRAMS**

UML supports twelve diagramming methods including structural (class, object, component, deployment), behavioral (case, sequence, activity, collaboration, state chart) and model management (incorporating packages, subsystems and models).

The Unified Modeling Language or UML is a mostly graphical modeling language that is used to express designs. It is important to understand that the UML describes a notation and not a process. It does not put forth a single method or process of design, but rather is a standardized tool that can be used in a design process.

**The Use Case Diagram**

In many design processes, the use case diagram is the first that designers will work with when starting a project. This diagram allows for the specification of high level user goals that the system must carry out. These goals are not necessarily tasks or actions, but can be more general required functionality of the system.

**Use Case**

More formally, a **use case** is made up of a set of **scenarios**. Each scenario is a sequence of steps that encompass an interaction between a user and a system. The use case brings scenarios together that accomplish a specific goal of the user.

A use case can be specified by textually describing the steps required and any alternative actions at each step. For example, the use case for searching a web for a keyword might be shown as:

1. Customer enters the keyword
2. Customer clicks the search button
3. The search is executed
4. The results are shown

Alternative: Search Failed
If the search fails at 3, then the user is redirected back to the search screen at step 1

In Visual Case, you can specify the steps of a use case in its description field. Simply right-click on a use case and select properties. You can then run a report and print or export the results to html or ascii text. Together, the report and the diagrams will include all of the details of the use case - their specific scenarios and the actors that carry them out.

**Actor**

The **use case diagram** allows a designer to graphically show these use cases and the actors that use them. An **actor** is a role that a user plays in the system. It is important to distinguish between a user and an actor (better thought of as a role). A user of the system may play several different roles through the course of his, her or its job (since an actor may be another system). Examples of actors are salesperson, manager, support person, and web store system. It is possible that the same person may be a sales person and also provide support. When creating a use case model, we are not concerned with the individuals, only the roles that they play.

**Associations**

On a use case diagram, associations are drawn between actors and use cases to show that an actor carries out a use case. A use case can be carried out by many actors and an actor may carry out many use cases.

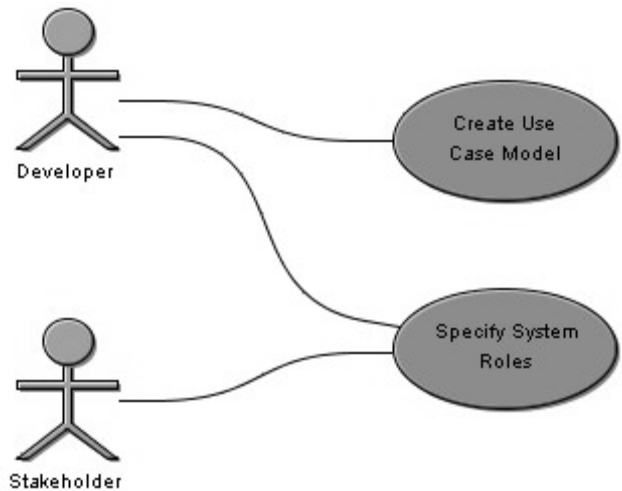

In the above diagram, the actors are shown as the green stick figure shapes on the left, the use cases are the blue ellipses, and the associations between them are represented by the connecting lines. The developer and the stakeholder both are responsible for specifying the system roles, but only the developer creates the model.

**Includes**

Use cases can also be related to each other with three different links. The diagram below shows the use of the includes link. Both *invoice purchase* and *online purchase* include the scenarios defined by *purchase valuation*. In general, the includes link is to avoid repetition of scenarios in multiple use cases.

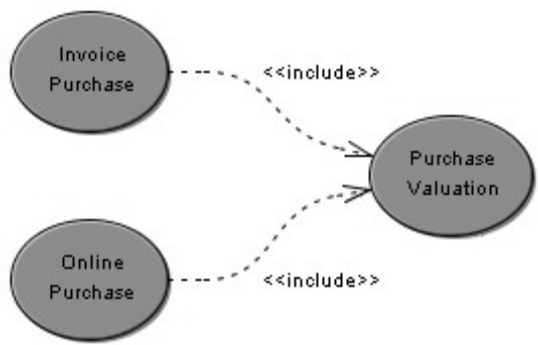

**Generalization**

When a use case describes a variation on another use case, use a generalization link. In the example below, the use case *limit exceeded* describes a situation in which the usual scenario of *online purchase* is not performed. Use cases that generalize another use case should only specify an alternative, even exceptional, scenario to the use case being generalized. The overall goal of the use cases should be the same.

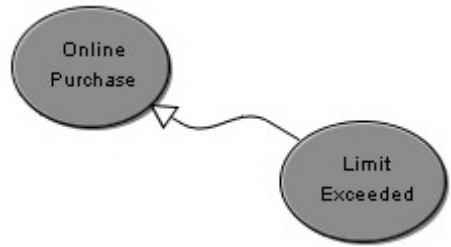

**Extends**

In some instances you want to describe a variation on behaviour in a more controlled form. In such instances you can define **extension points** in the extended use case. In the example below, *search by name* is said to extend *search* at the *name* extension point. The extends link is more controlled than the generalization link in that functionality can only be added at the extension points.

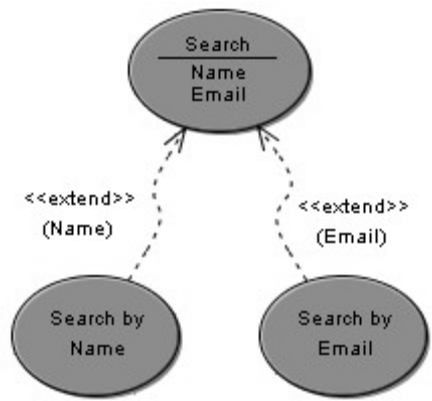

**Putting it all Together**

When starting a use case model, it is very important to keep it simple. Often it is easiest to first determine the actors of the system, and then flush out the use cases that they perform. Your use case diagrams can be as simple or complex as you wish, however simpler, less cluttered diagrams are easier to understand, and are often more powerful in capturing the tasks of the system.

In Visual Case, you can explode a use case into a new use case diagram. For example, the use case *online purchase* may require further specification as you move into the design. You can create a sub-diagram within any use case to help clarify and understand the tasks involved.

Remember that a use case represents a goal of a user, not an atomic programming operation. Your use case design should be simple and help to clarify the user's goals and expectations for the system.

**The Class Diagram**

The class diagram is core to object-oriented design. It describes the types of objects in the system and the static relationships between them.

**Packages**

Packages allow you to break up a large number of objects into related groupings. In many object oriented languages (such as Java), packages are used to provide scope and division to classes and interfaces. In the UML, packages serve a similar, but broader purpose.

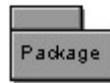

Any UML diagram in Visual Case can have packages on them. Each package can contain any type and any number of other UML diagrams, as well as interfaces and classes.

**Classes**

The core element of the class diagram is the class. In an object oriented system, classes are used to represent entities within the system; entities that often relate to real world objects.

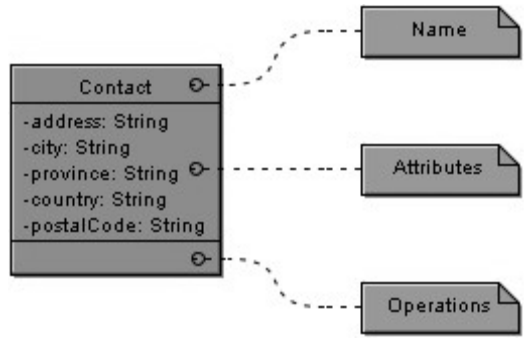

The *Contact* class above is an example of a simple class that stores location information.

Classes are divided into three sections:

**Top**: The **name**, **package** and **stereotype** are shown in the upper section of the class. In Visual Case, classes shown on a diagram that do not belong to the same package as the diagram are shown with their entire path name. You can optionally assign a stereotype to a class.

**Centre**: The centre section contains the attributes of the class.

**Bottom**: In the lower section are the **operations** that can be performed on the class.

On any Visual Case class diagram, you can optionally shown and hide both the attribute and operations sections of all the classes or individual classes. This is useful as often you will want your class diagrams to highlight specific constructs of your system that superfluous information only serves to clutter and confuse.

**Attributes**

An **attribute** is a property of a class. In the example above, we are told that a *Contact* has an address, a city, a province, a country and a postal code. It is generally understood that when implementing the class, functionality is provided to set and retrieve the information stored in attributes. Methods to set and retrieve attribute data are often called **accessor methods** (also **getting** and **setting** methods) and need not be shown in your model as they are usually inferred.

The format for attributes is:

*visibility name: type = defaultValue*

The visibility is as follows:

- **-** Private
- **+** Public
- **#** Protected
- **~** Package

In object oriented design, it is generally preferred to keep most attributes private as the accessor methods allow you to control access to the data. The most common exception to this preference are constants.

In addition to the name, visibility, datatype and default value, Visual Case allows you to specify the following properties for an attribute:

**Array**: you can set an attribute to be treated as an array of attributes; shown with square braces **[ ]** beside the name.

**Static**: attributes that are static only exist once for all instances of the class. In the example above, if we set *city* to be static, any time we used the *Contact* class the *city* attribute would always have the same value.

**Final:** if an attribute is declared final, it's value cannot be changed. The attribute is a constant.

**Operations**

The **operations** listed in a class represent the functions or tasks that can be performed on the data in the class.

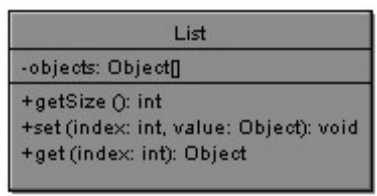

In the *List* class above, there is one attribute (a private array of Objects) and three operations.

The format for operations is:

*visibility name (parameters): type*

The format is very similar to that of the attribute except with the removal of a default value and the addition of parameters.

Parameters take the format:

*direction name: type = default value*

The direction can be one of *in*, *out*, *inout* or it can be unspecified.

In Visual Case you can show and hide the parameter list for a class or all classes on a diagram. If the list is hidden and an operation has parameters, three dots are shown (...) to indicate that parameters exist, but are hidden. Sometimes operations have numerous parameters that need not be shown all the time.

**Generalization**

The **generalization** link is used between two classes to show that a class incorporates all of the attributes and operations of another, but adds to them in some way.

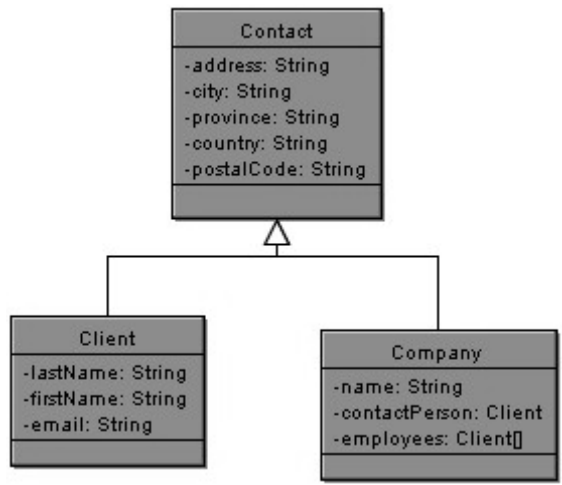

In the above diagram, we again see our *Contact* class, only now with two child classes. We can say that *Client* and *Company* **inherit**, **generalize** or **extend** *Contact*. In each of *Client* and *Company* all of the attributes in *Contact* (address, city, etc.) exist, but with more information added. In the above situation *Contact* is said to be the **superclass** of *Client* and *Company*.

When using a generalization link, the child classes have the option to **override** the operations in the parent class. That is, they can include an operation that is defined in the superclass, but define a new implementation for it.

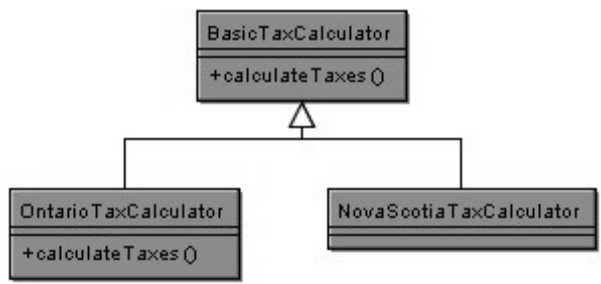

Above, *OntarioTaxCalculator* redefines or **overrides** the implementation of the method in BasicTaxCalculator. Essentially, the code is different but the operation is called the same way.

Sometimes you may want to force children to override methods in a parent class. In this case you can define the methods in the superclass as **abstract**. If a class has abstract operations, the class itself is considered abstract. Abstract methods and classes are shown in italics. Not all of the operations in an abstract class have to be abstract.

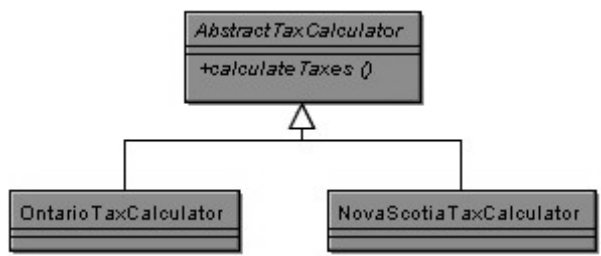

The abstract operation *calculateTaxes* in *AbstractTaxCalculator* must be implemented in the child classes OntarioTaxCalculator and NovaScotiaTaxCalculator. Since the operations must be implemented, it is not necessary to show them in the child classes, however you may if you choose. The key is to keep your diagrams as clear as possible. In the above instance the diagram is simple and the meaning clear, however with multiple levels of inheritance and more attributes and operations, you may wish to specify all of the methods that are overriden.

**Interfaces**

Many object oriented programming languages do not allow for multiple inheritance. The **interface** is used to solve the limitations posed by this. For example, in the earlier class diagram *Client* and *Company* both generalize *Contact* but one or the other child classes may have something in common with a third class that we do not want to duplicate in multiple classes.

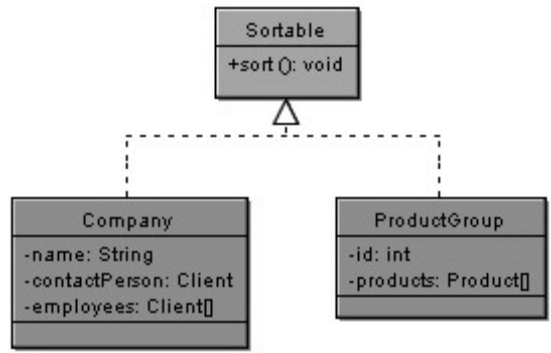

The interface *Sortable*, is used in the above example to show that both *Company* and *Product* implement the *sort* operation. We can say that *Company* and *Product* **implement** *Sortable* or that they are *Sortable*. Because Product already generalizes *Contact*, we could not also allow it to generalize *Sortable*. Instead, we made *Sortable* an interface and added a **realization** link to show the implementation.

Interfaces are very similar to abstract classes with the exception that they do not have any attributes. As well, unlike a class, all of the operations in an interface have no implementation. The child classes *Company* and *Product* are forced to implement the *sort* operation in its entirety.

**Associations**

Classes can also contain references to each other. The *Company* class has two attributes that reference the Client class.

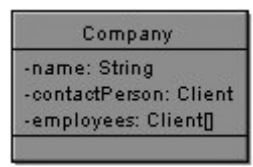

Although this is perfectly correct, it is sometimes more expressive to show the attributes as **associations**.

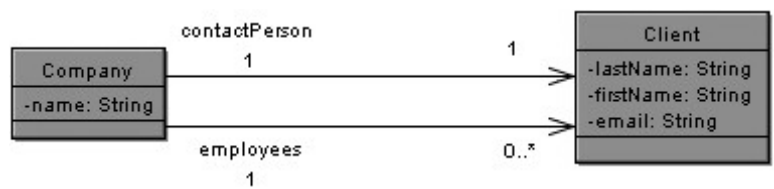

The above two associations have the same meaning as the attributes in the old version of the *Contact* class.

The first association (the top one) represents the old *contactPerson* attribute. There is one contact person in a single *Company*. The **multiplicity** of the association is one to one meaning that for every *Company* there is one and only one *contactPerson* and for each *contactPerson* there is one *Company*. In the bottom association there are zero or many employees for each company. Multiplicities can be anything you specify. Some examples are shown:

| | |
|---|---|
| **0** | zero |
| **1** | one |
| **1..*** | one or many |
| **1..2, 10..*** | one, two or ten and above but **not** three through nine |

The arrows at the end of the associations represent their **navigability.** In the above examples, the *Company* references *Clients*, but the *Client* class does not have any knowledge of the *Company*. You can set the navigability on either, neither or both ends of your associations. If there is no navigability shown then the navigability is unspecified.

**Aggregation and Composition**

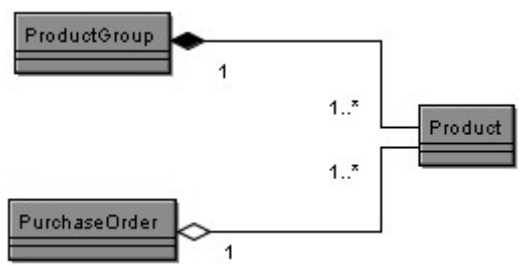

The above example shows an **aggregation** association and a **composition** association.

The **composition** association is represented by the solid diamond. It is said that *ProductGroup* is **composed** of *Products*. This means that if a *ProductGroup* is destroyed, the *Products* within the group are destroyed as well.

The **aggregation** association is represented by the hollow diamond. *PurchaseOrder* is an **aggregate** of *Products*. If a *PurchaseOrder* is destroyed, the *Products* still exist.

If you have trouble remembering the difference between composition and aggregation, just think of the alphabet. Composition means destroy and the letters 'c' and 'd' are next to each other.

**Dependencies**

A **dependency** exists between two elements if changes to one will affect the other. If for example, a class calls an operation in another class, then a dependency exists between the two. If you change the operation, than the dependent class will have to change as well. When designing your system, the goal is to minimize dependencies.

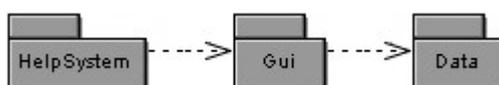

To help clarify the dependencies in your design, you may wish to draw a **Package Diagram**. A package diagram is essentially a class diagram with only packages and dependencies showing. Dependencies can exist between any components in the UML however at the highest level, dependencies will exist between packages. Within a package, the dependencies may be too numerous to specify. That is not to say that numerous dependencies are okay. Even within a package you want to limit the dependencies, however between packages in particular you should be strict about the number of dependencies that exist. In general, the fewer the dependencies the more **scaleable** and **maintainable** your system will be.

**Putting it all Together**

Class diagrams really are the core of most object oriented design so you will likely find yourself using them all the time. Fortunately class diagrams closely relate to the most object oriented languages, so the basics (classes, operations, attributes, generalizations, etc.) should be fairly easy to grasp. Start with what you know and then move on.

The most important thing about design is to not let it bog you down with detail. It is better to have a few clear diagrams than many, overly complex diagrams. Previously we saw the *AbstractTaxCalculator* that was generalized by *OntarioTaxCalculator* and *NovaScotiaTaxCalculator*. If we tried to create a diagram with all ten Canadian provinces and the three territories we would have a huge complex mess. If we were designing a tax system for the United States and we tried to show all of the states, we would be in even more trouble. It is more clear, and just as expressive to show two or three child classes and add a note to the diagram the explains that the other provinces and territories are implemented in the same way.

Keeping your designs simple will allow you to be more productive as well as making your designs far more understandable and useable. Also, as the system is implemented and upgraded, you'll want to keep your design in synch with your implementation. This will be far easier with a simple design of the key concepts of the system.

**Interaction Diagrams - Sequence and Collaboration**

Once the use cases are specified, and some of the core objects in the system are prototyped on class diagrams, we can start designing the dynamic behaviour of the system.

Recall from the Use Case page of this tutorial, that a use case *encompasses an interaction between a user and a system*. Typically, an **interaction diagram** captures the behaviour of a single case by showing the collaboration of the objects in the system to accomplish the task. These diagrams show objects in the system and the messages that are passed between them.

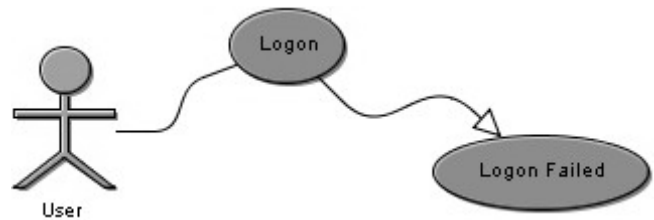

Let's start with the simple example above: a user logging onto the system. The *Logon* use case can be specified by the following step:

1. Logon dialog is shown
2. User enters user name and password
3. User clicks on OK or presses the enter key
4. The user name and password are checked and approved
5. The user is allowed into the system

Alternative: Logon Failed - if at step 4 the user name and password are not approved, allow the user to try again

Now that we have a simple Use Case to work with, we can specify some of the classes involved in the interaction.

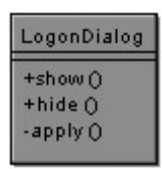

The *LogonDialog* has public methods to show and hide the window, and a private method that is called when the user presses the ok button or clicks enter. For our example (and indeed most cases) you need not specify the interface elements of the dialog.

Our design also includes a *LogonManager* class that will include one method that returns true if the logon is successful, false if it is not.

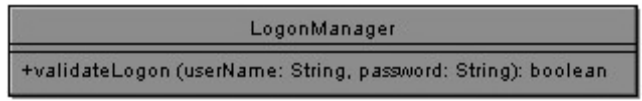

A *DatabaseAccess* class will allow us to run queries against our database. We can pass a query string and a *ResultSet* of data will be returned.

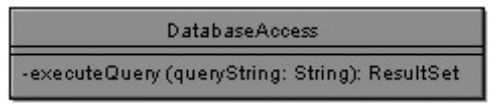

Now that we have prototyped the classes involved in our interaction, we can begin to make our interaction diagrams.

**Instances and Messages**

Interaction diagrams are composed mainly of instances and messages. An **instance** is said to be the realization of a class, that is if we have a class *Doctor*, than the instances are *Dr. Jones*, *Dr. Smith*, etc.. In an object oriented application, instances are what exist when you **instantiate** a class (create a new variable with the class as its datatype).

In the UML, instances are represented as rectangles with a single label formatted as:

<div style="text-align:center">instanceName: datatype</div>

You can choose to name the instance or not, but the datatype should always be specified.

Below the name, you can also list the attributes and their values. In Visual Case, you can map attributes from your class and enter new values specific to that instance. Attributes need only be shown when they are important and you don't have to specify and show all of the attributes of a class.

**Messages** represent operation calls. That is, if an instance calls an operation in itself or another class, a message is passed. Also, upon the completion of the operation a return message is sent back to the instance that initiated the call.

The format for message labels is:

<div style="text-align:center">Sequence Iteration [Guard] : name (parameters)</div>

**Sequence:** represents the order in which the message is called. The sequence is redundant on sequence diagrams, but required on collaboration diagrams

**Iteration:** an asterix (∗) is shown to represent iteration if the message is called repeatedly

**Guard:** an optional boolean expression (the result is either true or false) that determines if the message is called

**Name:** represents the operation being called

**Parameters:** represent the parameters on the operation being called

**Sequence Diagrams**

There are two types of Interaction Diagrams - **Sequence** and **Collaboration**.  Sequence diagrams emphasize the order in which things happen, while collaboration diagrams give more flexibility in their layout.  You can use whichever you prefer when drawing interactions, as both show the same information.

An example sequence diagram for our logon collaboration is shown:

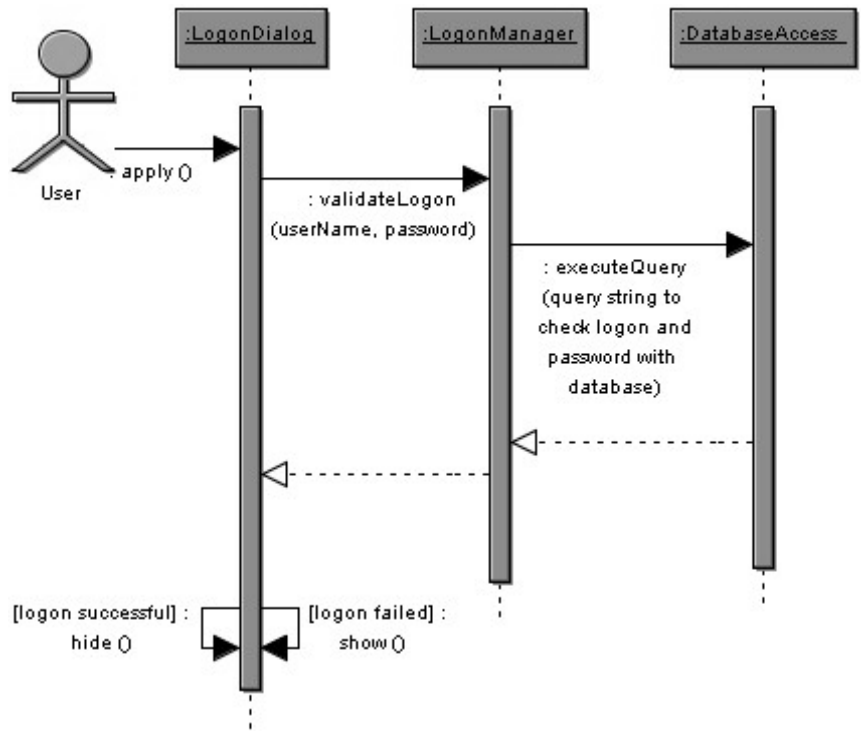

Things to Note:

- The flow of time is shown from top to bottom, that is messages higher on the diagram happen before those lower down
- The blue boxes are **instances** of the represented classes, and the vertical bars below are **timelines**
- The arrows (links) are **messages** - operation calls and returns from operations
- The hide and show messages use **guards** to determine which to call.  Guards are always shown in square braces **[ ]** and represent constraints on the message (the message is sent only if the constraint is satisfied)
- The messages are labelled with the operation being called and parameters are shown.  You can choose to enter the parameters or not - this is dependent upon their importance to the collaboration being shown
- The sequence numbers are not shown on the messages as the sequence is intrinsic to the diagram

**Asynchronous Messages**

You can specify a message as **asynchronous** if processing can continue while the message is being executed.  In the example below, the asynchronous call does not block processing for the regular call right below.  This is useful if the operation being called is run remotely, or in another thread.

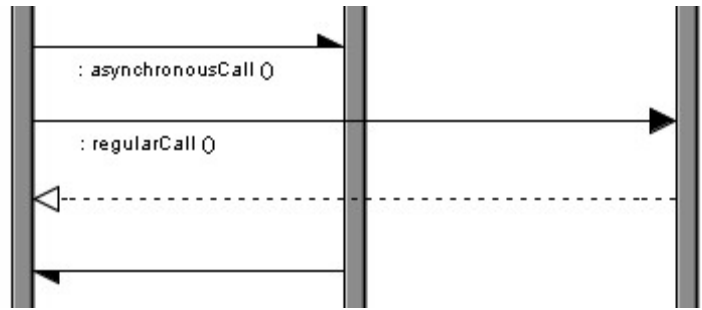

**Collaboration Diagrams**

Collaborations are more complex to follow than sequence diagrams, but they do provide the added benefit of more flexibility in terms of spatial layout.

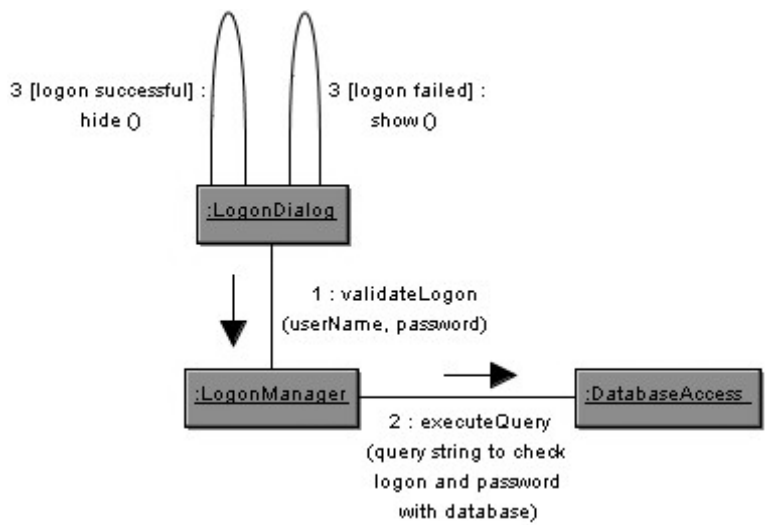

Above is our logon interaction shown as a collaboration diagram. Notice that each message is numbered in sequence, because it is not obvious from the diagram, the order of the messages.

**Lollipop Interfaces**

Another advantage over the sequence diagram is that collaboration diagrams allow you to show lollipop interfaces.

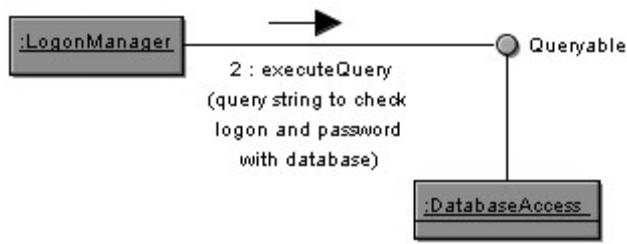

Suppose that our *DatabaseAccess* class implemented an interface called *Queryable*. If the logon manager only has access to the interface, we can show that the message is called through the interface by including a **lollipop interface** on the diagram. The stick of the lollipop indicates that the class *DatabaseAccess* realizes *Queryable*.

**Putting it all Together**

Using interaction diagrams, we can clarify the sequence of operation calls among objects used to complete a single use case. When drawing these diagrams, try to keep them as clear and simple as possible. Sequence diagrams are easy to read and follow, as they enforce a standard layout on the diagram. Collaborations have the added advantage of interfaces and freedom of layout, but can be difficult to follow, understand and create.

It's also important not to confuse interaction diagrams with state and activity diagrams. Interaction diagrams are used to diagram a single use case. When you want to examine the behaviour of a single instance over time use a state diagram, and if you want to look at the behaviour of the system over time use an activity diagram.

**Activity and State Diagrams**

Previously we saw how interaction diagrams demonstrate the behaviour of several objects when executing a single use case. When you want to show the sequence of events on a broader scale use activity and state diagrams.

**Activity Diagram**

An **activity** is the execution of a task whether it be a physical activity or the execution of code. Simply put, the **activity diagram** shows the sequence of activities. Like the simple flow chart, activity diagrams have support for conditional behaviour, but has added support for parallel execution as well.

**Start:** each activity diagram has one start (above) at which the sequence of actions begins.

**End:** each activity diagram has one finish at which the sequence of actions ends

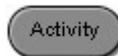

**Activity:** activities are connected together by transitions. **Transitions** are directed arrows flowing from the previous activity to the next activity. They are optionally accompanied by a textual label of the form:

**[guard] label**

The **guard** is a conditional expression that when true indicates that the transition is taken. The **label** is also optional and is freeform.

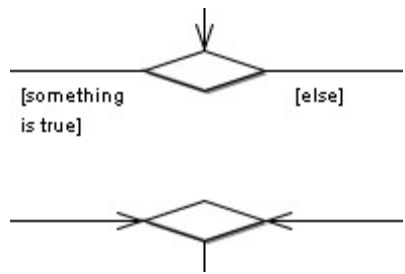

To show **conditional behaviour** use a branch and a merge. The top diamond is a **branch** and has only one transition flowing into it and any number of mutually exclusive transitions flowing out. That is, the guards on the outgoing transitions must resolve themselves so that only one is followed. The **merge** is used to end the conditional behaviour. There can be any number of incoming, and only one outgoing transition.

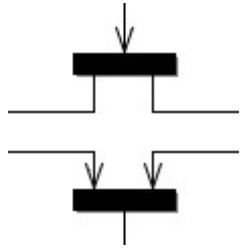

To show **parallel behaviour** use a fork and a join. The **fork** (top) has one transition entering and any number of transitions exiting, all of which will be taken. The **join** (bottom) represents the end of the parallel behaviour and has any number of transitions entering, and only one leaving.

**State Diagram**

The **state diagram** shows the change of an object through time. Based upon events that occur, the state diagram shows how the object changes from start to finish.

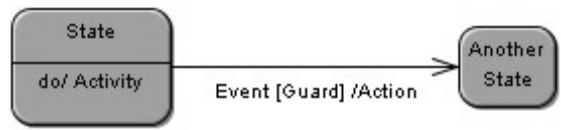

**States** are represented as a rounded rectangle with the name of the state shown. Optionally you can include an **activity** that represents a longer running task during that state.

Connecting states together are **transitions**. These represent the **events** that cause the object to change from one state to another. The **guard** clause of the label is again mutually exclusive and must resolve itself to be either **true** or **false**. **Actions** represent tasks that run causing the transitions.

Actions are different from activities in that actions cannot be interrupted, while an activity can be interrupted by an incoming event. Both ultimately represent an operation on the object being studied. For example, an operation that sets an attribute would be considered an action, while a long calculation might be an activity. The specific separation between the two depends on the object and the system being studied.

Like activity diagrams, state diagrams have one **start** and one **end** from at which the state transitions start and end respectively.

**Putting it all Together**

Activity diagrams are used to show workflow in parallel and conditionally. They are useful when working out the order and concurrency of a sequential algorithm, when analyzing the steps in a business process and when working with threads.

State diagrams show the change of an object over time and are useful when an object exhibits interesting or unusual behaviour - such as that of a user interface component.

As always, use these diagrams only when they serve a purpose. Don't feel that you have to draw a state diagram for every object in your system and an activity diagram for every process. Use them where they add to your design. You may not even include these diagrams in your design, and your work may still be complete and useful. The purpose of Visual Case and of the diagrams is to help you do your job, when the diagram becomes too complicated and pedantic, you and those working with you will lose focus on the task at hand.

**Implementation Diagrams - Component and Deployment**

So far, we've seen how to diagram the tasks that your system will perform, the details of the classes in your system, and the dynamic behaviour of the system - but what about the big picture? The two types of implementation diagrams provide the solution. With the **deployment** diagram, you can show how the components of your system are physically related, and with the **component** diagram, you can show the components in the system are organized.

You can combine the two diagrams if you choose:

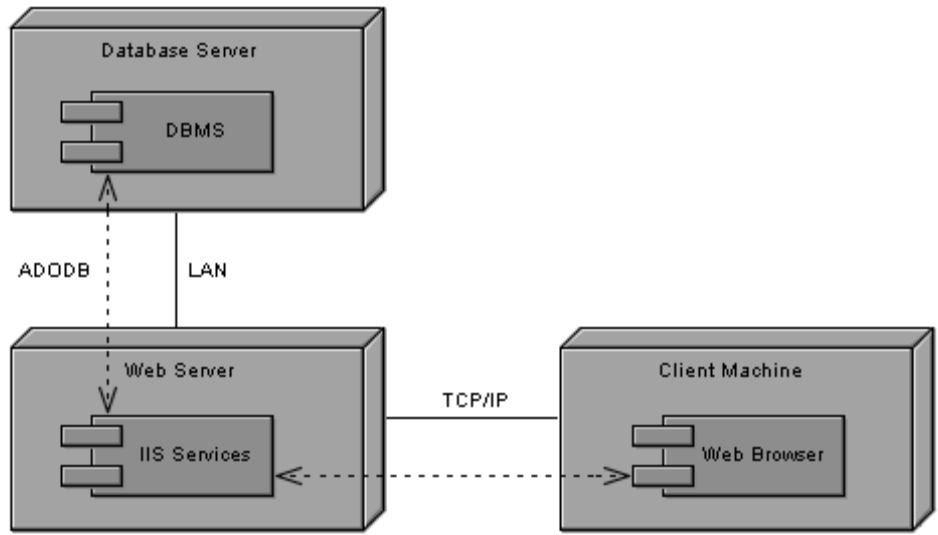

Above, the nodes are shown in green and the components in maroon. The **nodes** represent something upon which a component can run, and **components** are units of software.

On the diagram, nodes are connected with **connections** that show the physical path of information in the system. Components are connected with directed dashed lines that represent the **communication** between components. You can also use **lollipop interfaces** on components to show that communication is through an interface.

The important point to note here is that UML is a 'language' for specifying and not a method or procedure. The UML is used to define a software system; to detail the artifacts in the system, to document and construct - it is the language that the blueprint is written in. The UML may be used in a variety of ways to support a software development methodology (such as the Rational Unified Process) - but in itself it does not specify that methodology or process.

UML defines the notation and semantics for the following domains:

- The User Interaction or Use Case Model - describes the boundary and interaction between the system and users. Corresponds in some respects to a requirements model.

- The Interaction or Communication Model - describes how objects in the system will interact with each other to get work done.

- The State or Dynamic Model - State charts describe the states or conditions that classes assume over time. Activity graphs describe the workflows the system will implement.

- The Logical or Class Model - describes the classes and objects that will make up the system.

- The Physical Component Model - describes the software (and sometimes hardware components) that make up the system.

- The Physical Deployment Model - describes the physical architecture and the deployment of components on that hardware architecture.

**CONCLUSION**

The UML is a language in that it specifies a vocabulary and rules for combining words in that vocabulary. It is a modeling language because its vocabulary and rules focus on the conceptual and
physical representation of a system. The Unified Modeling Language (UML) has quickly become the de-facto standard for building Object-Oriented software. The important point to note here is that UML is a 'language' for specifying and not a method or procedure. The UML is used to define a software system; to detail the artifacts in the system, to document and construct - it is the language that the blueprint is written in. The UML may be used in a variety of ways to support a software development methodology (such as the Rational Unified Process) - but in itself it does not specify that methodology or process.